\documentclass[12pt]{article}
\usepackage{fullpage,latexsym,epsfig,amsfonts}

\renewcommand{\phi}{\varphi}

\newtheorem{definition}{Definition}[section]

\newtheorem{lemma}[definition]{Lemma}
\newtheorem{theorem}[definition]{Theorem}

\newenvironment{proof}{\noindent{\it Proof.}}{\hfill$\Box$}

\title{A note on the classical lower bound for a quantum walk algorithm}

\author{Stephen A. Fenner\thanks{Computer Science and Engineering
Department, University of South Carolina, Columbia, SC 29208 USA.
Email {\tt \{fenner|zhang29\}@cse.sc.edu}.  This work was supported in
part by the National Security Agency (NSA) and Advanced Research and
Development Activity (ARDA) under Army Research Office (ARO) contract
number DAAD~190210048.}
\and
Yong Zhang\footnotemark[1]
}

\date{June 27, 2003}

\begin{document}

\maketitle

\begin{abstract}
A recent paper on quantum walks by Childs et al.\
\cite{CCDFGS:quantum-walk} provides an example of a black-box problem
for which there is a quantum algorithm with exponential speedup over
the best classical randomized algorithm for the problem, but where the
quantum algorithm does not involve any use of the quantum Fourier
transform.  They give an exponential lower bound for a classical
randomized algorithm solving the black-box graph traversal problem
defined in their paper.  In this note we give an improved lower bound
for this problem via a straightforward and more complete analysis.
\end{abstract}

\section{Introduction}

Almost all the quantum algorithms with exponential speedup over their
best known classical counterparts use some type of quantum Fourier
transform---even for problems (such as shifted quadratic characters)
which are not obvious instances of the Hidden Subgroup problem.
However, a recent paper on quantum walks \cite{CCDFGS:quantum-walk}
does not fall into this category.  They demonstrate that exponential
speedup can be achieved by a different algorithmic technique, the
quantum walk. They give a polynomial-time quantum algorithm to solve a
black-box graph traversal problem, based on a continuous time quantum
walk.  Their paper contains three major results: the continuous time
quantum walk algorithm itself, its implementation by a circuit with
(discrete) quantum gates, and an exponential lower bound for solving
the problem classically.  We refer to their paper for a detailed
description.  Here we concentrate on improving their classical lower
bound.

They showed that any classical algorithm solving the problem of
traversing $G_n'$ (see Figure 2 in \cite{CCDFGS:quantum-walk})
requires exponential time:

\begin{theorem}[\protect{\cite[Theorem 6]{CCDFGS:quantum-walk}}]
Any classical algorithm that makes at most $2^{n/6}$ queries to the oracle
finds the \rm{EXIT} with probability at most $4\cdot 2^{-n/6}$.
\end{theorem}

In this note we use a more straightforward approach to analyze this
problem.  Our analysis also considers some cases Childs et al.\
neglected.  Furthermore, we obtain an improved lower bound, as stated
in the following theorem.

\begin{theorem}
Any classical algorithm that makes at most $2^{n/3}$ queries to the
oracle finds the \rm{EXIT} with probability at most $O(n\cdot
2^{-n/3})$.
\end{theorem}

\section{An improved analysis on the classical lower bound}

In this section we generally refer to the set-up in Childs et al.\
\cite{CCDFGS:quantum-walk}.  Our improvement essentially comes from
the calculation of the probability that a classical randomized
algrorithm wins Game 4.  We do not need to consider subtrees of $G_n'$
of height $n/2$.

\begin{lemma}[cf.\ \protect{\cite[Lemma 10]{CCDFGS:quantum-walk}}]
For any rooted tree $T$ of at most $2^{n/3}$ vertices,
\[ \max_{T}E_G[P^G(T)] = O(n\cdot 2^{-n/3}). \]
\end{lemma}

\begin{proof}
Let $T$ be a tree with $t$ vertices, $t\leq 2^{n/3}$, with image $\pi
(T)$ in $G_n'$ under the random embedding $\pi$.  For any nonroot node
$u \in T$, let $p(u)\in T$ be the parent of $u$.

We assume that in $G_n'$, the ENTRANCE is at level $0$ and the EXIT is
at level $2n+1$, and the middle layer is between levels $n$ and $n+1$.
Thus both binary trees have height $n$.  To reach the EXIT from the
column $n+1$, $\pi$ has to move right $n$ times in a row, which has
probability $2^{-n}$. Since there are at most $t$ tries on each path
of $T$, and there are at most $t$ such paths, the probability of
finding the EXIT $\Pr[\mbox{$T$ exits on $G_n'$}]$ is bounded by
$t^2\cdot 2^{-n}$.

Now assume that $\pi(T)$ does not exit.  It is easy to see that 
\begin{eqnarray} \label{bigsum}
\Pr[\mbox{$\pi$ is improper}] & = & \sum_{a,b\in T}\Pr[\pi(a)=\pi(b)
 \; \&\; \pi(p(a)) \neq \pi(p(b))] \\
 & = & \sum_{a,b\in T}\sum_{u\in G_n'}\Pr[\pi(a)=\pi(b)=u \; \&\;
 \pi(p(a)) \neq \pi(p(b))].
\end{eqnarray}
We calculate $\Pr[\pi(a)=\pi(b)=u]$ first for fixed $a,b\in T$, and
$u\in G_n'$.  Let the \emph{height} of $u$ be the distance $h$ from $u$ to
a leaf of the subtree that $u$ is in.  That is, if $u$ is at level
$\ell \leq n$, then $h = n-\ell$, and if $u$ is at level
$\ell \geq n+1$, then $h = \ell - n - 1$.

Look at the two paths $p_a$ and $p_b$ in $T$ from the root to $a$ and
to $b$, respectively.  Either both paths go through the middle layer
of $G_n'$ or only one path goes through the middle layer (at least one
path must go through the middle layer).  Assume first that both paths
go through the middle layer.

Assume WLOG that $p_b$ goes through the middle layer at least as many
times as $p_a$.  From the last time $p_a$ goes through the middle
layer, in order to reach $u$ it must first reach some (possibly
improper) ancester $v$ of $u$.  If $v$ has height $h'$, then the
probability of $p_a$ getting to $v$ from just before the last middle
layer is $2^{-h'} \cdot 2^{h'}/(2^n - t) = 1/(2^n - t)$. Once reaching
$v$, the probability of then going to $u$ is then $2^{h-h'}$.  So the
probability of $p_a$ reaching $u$ is at most
\[ \sum_{h'=h}^n \frac{2^{h-h'}}{2^n - t} < \frac{2}{2^n - t}. \]
By our WLOG assumption, there is a final subpath $p_b'$ of $p_b$ not
on $p_a$ that goes through the middle layer before reaching $u$.  By
the same analysis, we get that $p_b'$ reaches $u$ with probability
less than $2/(2^n - t)$.  The probabilities of both paths reaching $u$
is thus less than $4/(2^n - t)^2$.  (The two paths $p_a$ and $p_b'$
are almost independent, but not quite; the dependence only decreases
the true probability from the product above, however.  With a finer
analysis, one can reduce the probability to $3/[2(2^n - t)^2]$.)

Now assume that only one path, say $p_b$ goes through the middle
layer, and that $\pi(a) = \pi(b) = u$.  Then $a$ is an ancestor of $b$
with depth $\leq n$ in $T$, and $\pi(a)$ is the leftmost point of the
cycle, lying in the left half of $G_n'$.  The probability that $p_a$
ends with $u$ is thus at most $2^{h-n}$, where $h$ is the height of
$u$.  Let $p_b'$ be defined as before.  Then by an analysis similar to
that above, $p_b'$ ends in $u$ with probability $1/(2^n - t)$ (note
that $p_b'$ approaches $u$ from the right, i.e., it cannot go through
a proper ancestor of $u$ first).  By (almost) independence, we then
have in this case,
\[ \Pr[\pi(a)=\pi(b) \; \&\; \pi(p(a)) \neq \pi(p(b))] \leq
\frac{2^{h-n}}{2^n - t}. \]
Summing this probability over $u$ in the left half of $G_n'$, we get
\[ \sum_{u} \frac{2^{h-n}}{2^n - t} = \sum_{h=0}^n 2^{n-h} \cdot
\frac{2^{h-n}}{2^n - t} = \frac{n+1}{2^n - t}. \]

We can now easily compute the sum (\ref{bigsum}).  We get that $\pi$
is improper with probability less than
\begin{eqnarray*}
\lefteqn{\sum_{a,b\in T} \left( 2^{n+2}\cdot\frac{4}{(2^n - t)^2} +
\frac{n+1}{2^n - t} \right)} \\
& \leq & \frac{t^2}{2^n - t}\left(\frac{2^{n+2}}{2^n - t} + n + 1
\right) \\
& = & O(n\cdot 2^{n/3})
\end{eqnarray*}
if $t \leq 2^{n/3}$.
\end{proof}


\newcommand{\etalchar}[1]{$^{#1}$}

\end{document}